\documentclass[journal = jctcce, manuscript=article]{achemso}
\usepackage{dcolumn}
\usepackage{bm}
\usepackage{amsmath}
\usepackage{rotating}
\usepackage{graphicx}
\usepackage{gensymb}
\usepackage{xcolor}
\usepackage{makecell}
\usepackage{booktabs}
\usepackage{lscape}
\usepackage{longtable}
\author{Pranesh Raghavendran}
\affiliation{Department of Chemistry and Biochemistry, Butler University, Indianapolis, IN 46208 USA}
\author{Gabriela E. Campbell}
\affiliation{Department of Chemistry and Biochemistry, Butler University, Indianapolis, IN 46208 USA}
\author{Erik P. Hoy}
\affiliation{Department of Chemistry, Rowan University, Glassboro, NJ USA}
\author{Andrew M. Sand}
\affiliation{Department of Chemistry and Biochemistry, Butler University, Indianapolis, IN 46208 USA}
\email{amsand@butler.edu}

\title{Exploring electron transport in carbon wire systems using a wave function-based multiconfigurational non-equilibrium Green's function approach}
\date{\today}

\begin{document}

\maketitle
\begin{abstract}
 
One-dimensional carbon wire materials are an interesting class of materials which have potentially useful applications in sub-nanoscale junction technology due to their unique quantum transport properties.  In this work, we apply a non-equilibrium Green's function approach based on multiconfiguration pair-density functional theory (NEGF-MCPDFT).  NEGF-MCPDFT is a wave-function-based non-periodic framework to study electron transport in systems exhibiting strong correlation.  The advantage of this approach is multiconfigurational wave functions are used which offer enhanced abilities to describe strong electron correlation beyond single-determinant approaches.  MC-PDFT necessitates the use of an active space, and a key step in such calculations is the determination of which orbitals to include.  In application of this theory to carbon wire junctions of various lengths, we find that NEGF-MCPDFT is capable of reproducing trends from recent experimental results, and we use our findings to further the development of best practices in active space design for transport calculations of this type.

\end{abstract}

\section{Introduction}
%background about nanowires, in particular carbon

As electronic devices approach the practical limits of miniaturization for traditional silicon technologies, attention, both theoretical and experimental, has turned towards atomic and molecular scale systems.  In addition to their small size, atomic or molecular junctions can exhibit quantum features that may be harnessed or exploited in future designs of junctions and devices.  One dimensional linear carbon chains or carbon wires represent one class of emerging technology~\cite{hirsch_era_2010,casari_carbon-atom_2016,casari_carbyne_2018,bryce_review_2021}, noted for their mechanical and electrical properties.

One-dimensional linear carbon wires, essentially consisting of a chain of $sp-$hybridized carbons, can adopt a variety of allotropes.  These include cumulenes (...=C=C=...) and polyynes/oligoyne (...-C$\equiv$C-...).  Cumulenes are generally thought to have more metallic electron transport properties relatively independent of length, while polyynes behave more as semiconductors with conductance decaying exponentially as a function of overall wire length~\cite{milani_carbon_2006,wang_oligoyne_2009}.  A bond-length alternation (BLA) parameter is widely used to characterize the structure of carbon wires, with a BLA=0 representing a perfect cumulene structure with identical C-C bond distances throughout and a $\mathrm{BLA} \neq 0$ representing some amount of bond distance difference between adjacent carbon bonds.  Experimentally, the connections which interface the carbon wire to the electrodes can influence the adoption of the carbon chain to more cumulene-like or polyyne/oligoyne-like structures ~\cite{schermann_dicyanopolyynes_1997,eisler_polyynes_2005,szafert_update_2006,chalifoux_synthesis_2008,shi_confined_2016,kaiser_sp-hybridized_2019,gao_loss_2020,morris_charge_2026}.  Several theoretical studies have also explored the transport properties of one-dimensional carbon systems~\cite{prasongkit_cumulene_2010,fang_electronic_2011,wang_first-principles_2012,al-backri_electronic_2014,kumar_effect_2015,sarbadhikary_magnetic_2018,garner_reverse_2018,ferreira_electronic_2020,xu_unusual_2019,zang_cumulene_2020,balakrishnan_dft_2020,garner_three_2020,balakrishnan_polyyne-metal_2021,mu_ab_2023,wang_exploring_2024} with a variety of different electrode models.  In general, the studies support the trend of cumulenic structures leading to better conductance properties.

%Importance of multireference correlation in the description of transport properties
The majority of simulations of charge transport in molecular electronics employ methods based on non-equilibrium Green's function theory~(NEGF), where DFT is commonly used to construct approximate Green's functions for the NEGF transport calculations~(NEGF-DFT).  The single-determinant approach of DFT, however, limits the ability of the approach to treat multireference effects.  With an ever-increasing set of systems that may exhibit unique electron transport properties~\cite{ren_collection_2024}, many are exploring the role that strong electron correlation may play in the emergence of these features~\cite{cossaboon_assessing_2024}.

Recently, there have been several alternative approaches presented which aim to more effectively treat strong correlation effects in electron transport~\cite{gandus_strongly_2026}.  Previously, our groups have introduced an NEGF theory that uses multiconfiguration pair-density functional theory (MC-PDFT)~\cite{mcpdft} in an approach we denote NEGF-MCPDFT~\cite{sand_multiconfiguration_2021}.  NEGF-MCPDFT allows for the inclusion of both static and dynamic electron correlations into a description of the transport properties of molecular junctions.  The underlying MC-PDFT is most commonly used with multiconfigurational reference wave functions which employ an active space formalism.  The selection of an active space is an important step in most multiconfigurational calculations.  Several studies~\cite{pulay_uhf_1988,veryazov_how_2011,stein_automated_2016,bao_automatic_2018,sayfutyarova_constructing_2019,khedkar_active_2019,kolodzeiski_automated_2023} have developed best practices for the determination of active orbitals, either requiring manual selection or the use of an automated procedure.  
Often, these procedures are developed to accommodate a certain task, such as minimization of ground-state energies, predicting accurate spectra, or generating orbitals suitable for quantum embedding calculations.
The rules governing best practices in active orbital selection for a multiconfigurational electron transport calculation, like NEGF-MCPDFT, are not fully explored.

%Model system to provide information on best practices for active space designs in NEGF-MCPDFT

In this work, we explore the electron transport properties of one-dimensional carbon systems using NEGF-MCPDFT.  We explore how calculated transport properties are affected by changing from a single reference to a multireference methodology, we explore how orbital optimization affects calculated transport properties, and from this, we develop some insights into appropriate active orbital selections for these calculations.

\section{Computational Details} 

All DFT, CASCI, and CASSCF reference calculations were performed using the OpenMolcas~\cite{li_manni_openmolcas_2023} software package.
Electron transport calculations employed the pyRUQT\cite{RUQT} package, which employs the Atomic Simulation Environment (ASE)~\cite{hjorth_larsen_atomic_2017} software package.  Visualization of the orbitals was performed with the Luscus software~\cite{kovacevic_luscus_2015}.

\begin{figure}
\begin{center}
\includegraphics[scale=1]{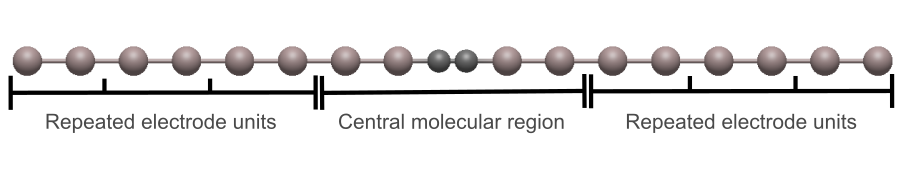}
\end{center}
\caption{Model of the system studied.}
\label{fig:syst}
\end{figure}

In an NEGF-MCPDFT calculation, the model system is divided into three parts: the left electrode, the central region, and the right electrode.  A representation of our system is given in Fig.~\ref{fig:syst}.  Aluminum is used as the electrode instead of more experimentally-relevant gold to manage the size of the system, allowing for larger electrode regions. The left and right electrodes consist of six aluminum atoms, equally spaced, 2.53 angstroms apart.  This bond distance was found using a periodic DFT calculation using SIESTA~\cite{soler_siesta_2002} (PBE, DZ basis set).  The central region consists of carbon wires of various lengths augmented on the left and right sides by an additional two aluminum electrode-like structure, allowing for more flexibility in the obtaining of an optimized geometry.  Structures were then optimized using KS-DFT with the PBE0 functional and the Dunning-Hay double zeta plus polarization (DZP) basis set~\cite{jr_gaussian_1989,dunning_gaussian_1977} for carbon atoms and the ECP-containing Stuttgart basis set~\cite{bergner_ab_1993} for aluminum atoms, freezing the bond distances between the six aluminum atoms in each electrode.  In total, eight aluminum atoms appear on the left and right sides of the carbon wires.  Additionally, we performed a non-periodic DFT calculation (PBE0/Stuttgart) with 16 aluminum atoms to approximate the Fermi level as the energetic midpoint between the HOMO and LUMO orbitals (-4.17 eV).

The delocalized PBE0 orbitals were used as the orbital guesses for multiconfigurational CASCI and CASSCF calculations.  We again used the DZP and Stuttgart basis sets for carbon and aluminum, respectively.  The tPBE pair-density functional was used~\cite{mcpdft}.  In CASCI, the orbitals remain unchanged and in CASSCF the orbitals are re-optimized.  A critical step in these CAS calculations is the determination of an appropriate active space.  For these systems, we targeted active orbitals near the HOMO/LUMO gap which contained significant orbital density in the electrode/central interface region.  Ideally, we want this active space to be small and manageable.  For systems with an even number of carbon atoms, a (6e,6o) active space was selected, and for systems with an odd number of carbon atoms, an (8e,8o) active space was selected.  The difference for the two active spaces was due to orbital degeneracies being present in the odd carbon cases; it would not be appropriate to break this degeneracy to obtain a size match with the even carbon cases.

\section{Results and Discussion}

\subsection{Structures}

\begin{figure}
\begin{center}
\includegraphics[scale=1]{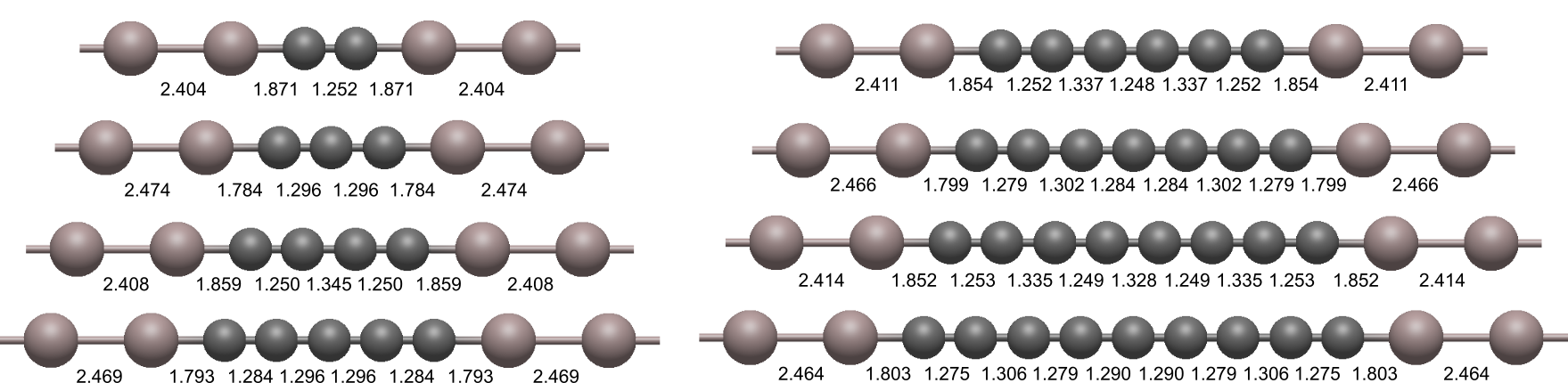}
\end{center}
\caption{Bond lengths of the optimized carbon wire junctions.  Units are given in \AA.}
\label{fig:structures}
\end{figure}

\begin{figure}
\begin{center}
\includegraphics[scale=0.5]{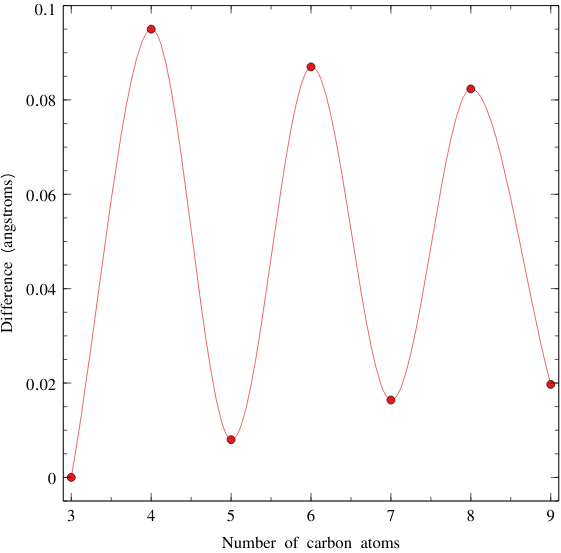}
\end{center}
\caption{Degree of BLA for carbon wires of different lengths.  The mean unsigned distance between two adjacent bonds is plotted on the vertical axis.}
\label{fig:BLA}
\end{figure}

The structures of the carbon wire systems studied are given in Fig.~\ref{fig:structures}.  There are distinct structural differences between systems with an even and odd number of carbon atoms.  
To more explicitly show the degree of BLA, a plot of the average distance between adjacent C-C bonds is given in Fig.~\ref{fig:BLA}.
Our results are in line with previous DFT studies on these systems~\cite{yang_linear_2007}.  End effects result in more acetylenic structures for even-carbon systems and cumulenic structures for odd-carbon systems, but the BLA becomes more uniform as the chain length increases.

\subsection{Orbitals}

\begin{figure}
\begin{center}
\includegraphics[scale=.6]{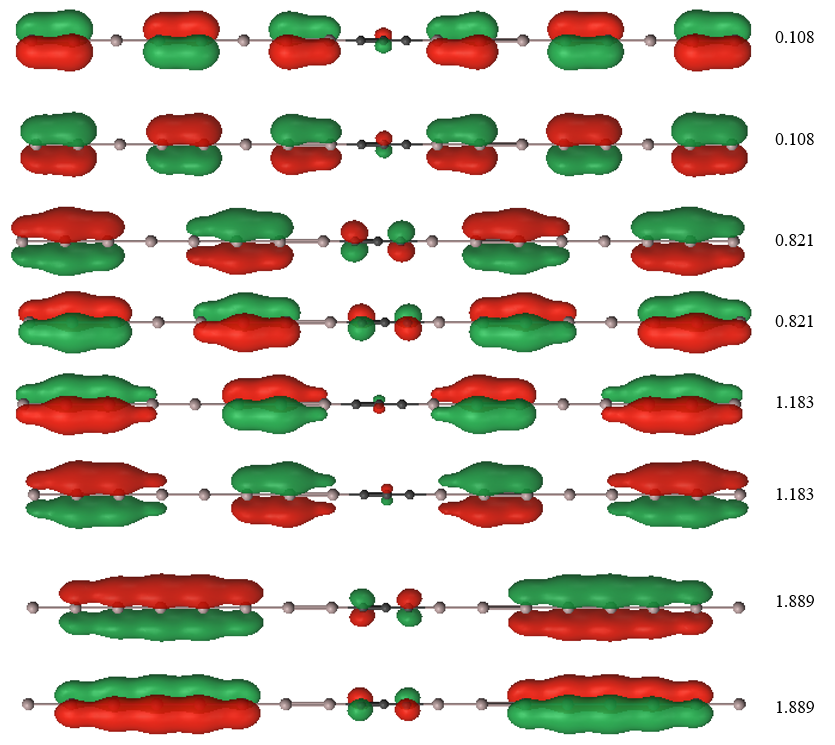}
\end{center}
\caption{PBE0 orbitals used to seed the CASCI calculation for the three-carbon case.  Resultant CASCI occupation numbers are given.}
\label{fig:3_orb}
\end{figure}

\begin{figure}
\begin{center}
\includegraphics[scale=.6]{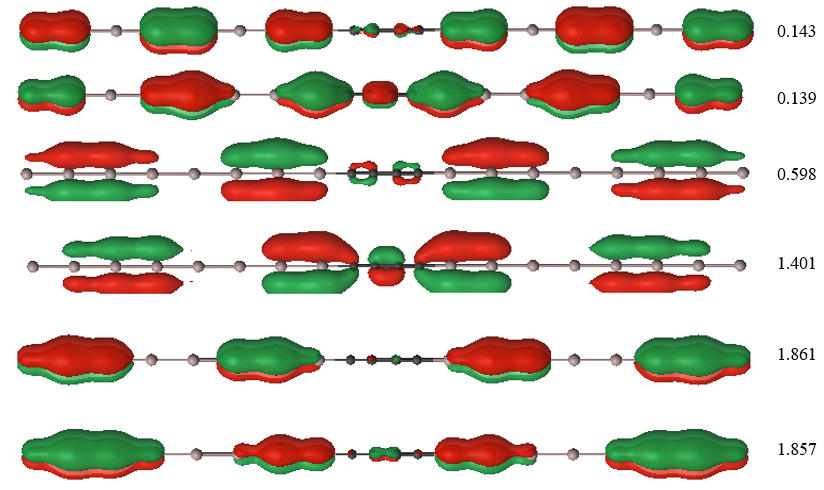}
\end{center}
\caption{PBE0 orbitals used to seed the CASCI calculation for the four-carbon case.  Resultant CASCI occupation numbers are given.}
\label{fig:4_orb}
\end{figure}

We have identified orbitals from the PBE0 calculations which we believe are the most critical for a good description of electron transport.  We present orbital images for a typical even-carbon case~(Fig.~\ref{fig:4_orb}) and a typical odd-carbon case~(Fig.~\ref{fig:3_orb}).  Again, we note that the active space sizes are different for the even and odd cases.  This is typically not ideal in comparative calculations, but here we believe it to be justified in order to not have a set of degenerate orbitals appear both inside and outside the active space.  These orbitals are deemed favorable because of their nearness in energy to the HOMO/LUMO levels, their significant orbital density in the interface region between the carbon chain and the electrodes, and their diffuse character, extending deep into both the central region and the electrode regions.  These orbitals were used for CASCI calculations, where the orbitals are not allowed to re-optimize, but the distribution of electrons between the active orbitals is optimized.

\begin{figure}
\begin{center}
\includegraphics[scale=.6]{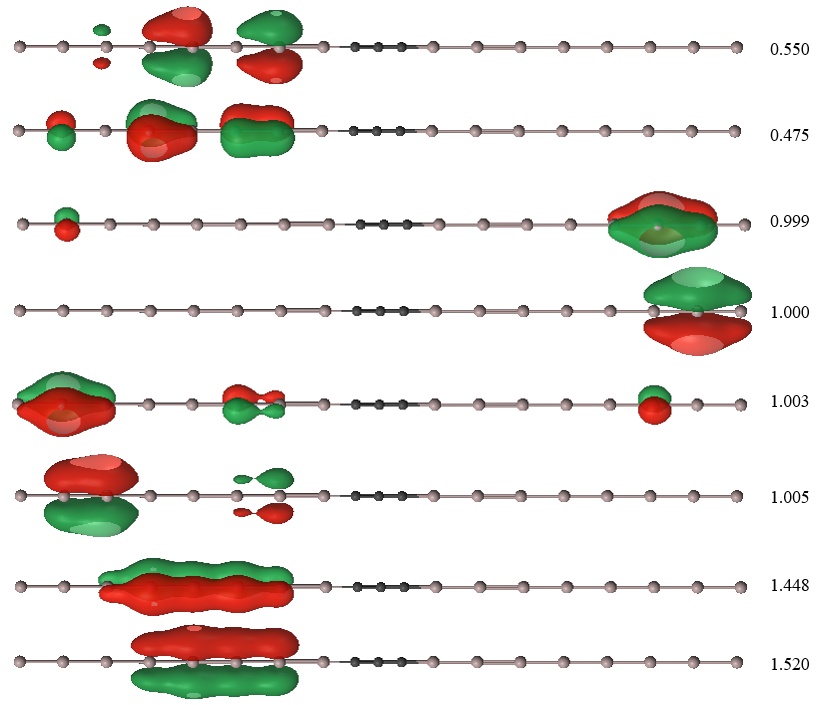}
\end{center}
\caption{CASSCF orbitals for the three-carbon case.  The initial guess orbitals were identical to the orbitals used in the CASCI calculation.}
\label{fig:SCF_3}
\end{figure}

\begin{figure}
\begin{center}
\includegraphics[scale=.6]{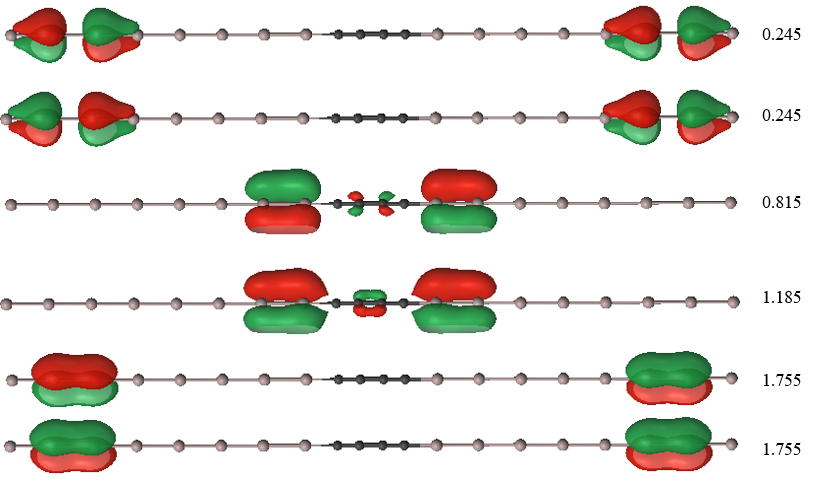}
\end{center}
\caption{CASSCF orbitals for the four-carbon case.  The initial guess orbitals were identical to the orbitals used in the CASCI calculation.}
\label{fig:SCF_4}
\end{figure}

Calculations were also performed with CASSCF, where the orbitals are re-optimized at the CASSCF level.  Prototypical examples for even- and odd-carbon systems are given in Figs.~\ref{fig:SCF_4}~and~\ref{fig:SCF_3}.  The CASSCF calculations were seeded with the identical orbitals used from the PBE0 calculations in the CASCI calculations.  We observe these orbitals have some significant potential issues for transport calculations (based on the results of a previous study~\cite{cossaboon_assessing_2024}): the orbitals tend to become very localized within the electrode regions and the orbitals are highly asymmetric.  There are two reasons for this.  First, the Al atoms in the electrode regions of the system are more multiconfigurational in character (as seen when comparing the CASCI versus CASSCF occupation numbers).  This results in orbital densities being highly encouraged to be more localized on the Al electrode atoms instead of the central or interface regions.  Second, this localization would now require a much larger active space to fully span the metallic character of the electrode regions (and also the interface/central region).  Expanding the active space by two or four orbitals does not, however, resolve the asymmetry issue.  A similar trend towards more electrode-focused orbitals was seen in other systems~\cite{cossaboon_assessing_2024}.  Because ultimately we aim to describe much larger systems with more complicated electrodes as opposed to a one-dimensional electrode, a large expansion of the active space is not a fruitful path forward.  Instead, to learn about the consequences of using less-than-ideal active spaces, we continue on and use this orbital set for the CASSCF transport results discussed in later sections.

\subsection{Electron transport with active space approaches}

\begin{figure}
\begin{center}
\includegraphics[scale=.5]{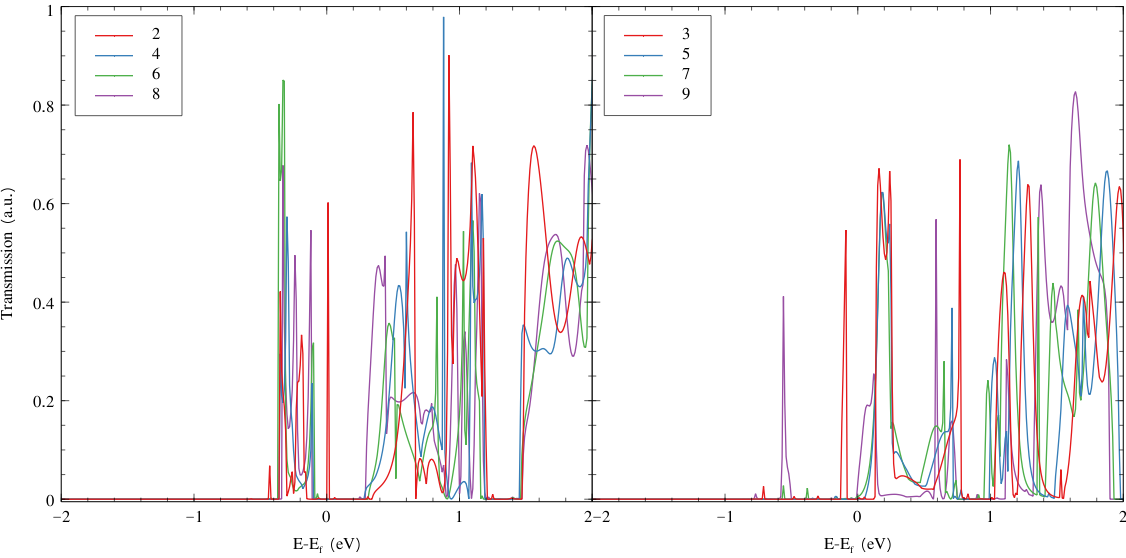}
\end{center}
\caption{Transmission functions for the carbon wire systems with NEGF-MCPDFT using the CASCI reference wave function.  Even and odd cases are grouped together.}
\label{fig:trans_CASCI}
\end{figure}

Transport properties were then calculated for these even- and odd-carbon systems.  Transmission functions for the systems using NEGF-MCPDFT with the CASCI reference are shown in Fig.~\ref{fig:trans_CASCI}.  All transmission is plotted relative to the approximate Fermi level, which is at 0~eV in all plots.  The functions are broken into two sets: even and odd numbers of carbon atoms.  There are similar patterns observed for both sets. 
For the even-length systems (of more polyyne character), the CASCI approach generally produces two  regions with significant transport near the Fermi level.  One set of transport signals is found between -0.5 to 0~eV and another set is found in the 0.2 to 1.2~eV region.  With the exception of the two-carbon chain (where a very narrow peak is very near 0~eV), we find a gap near the Fermi level.  This is indicative of semiconducting-like behavior, and this is consistent with experimental findings showing that polyynic wires are semiconductors.

For the odd-length systems (of more cumulenic character), all systems show narrow transmission peaks starting very near the Fermi level (0.0 to 0.6~eV) but with less grouping among the different length systems, and as the length of the chain increases, the overlap of the transmission signals with the Fermi level increases.  These results are indicative of conducting behavior and are consistent with previous experimental results.

\begin{figure}
\begin{center}
\includegraphics[scale=.5]{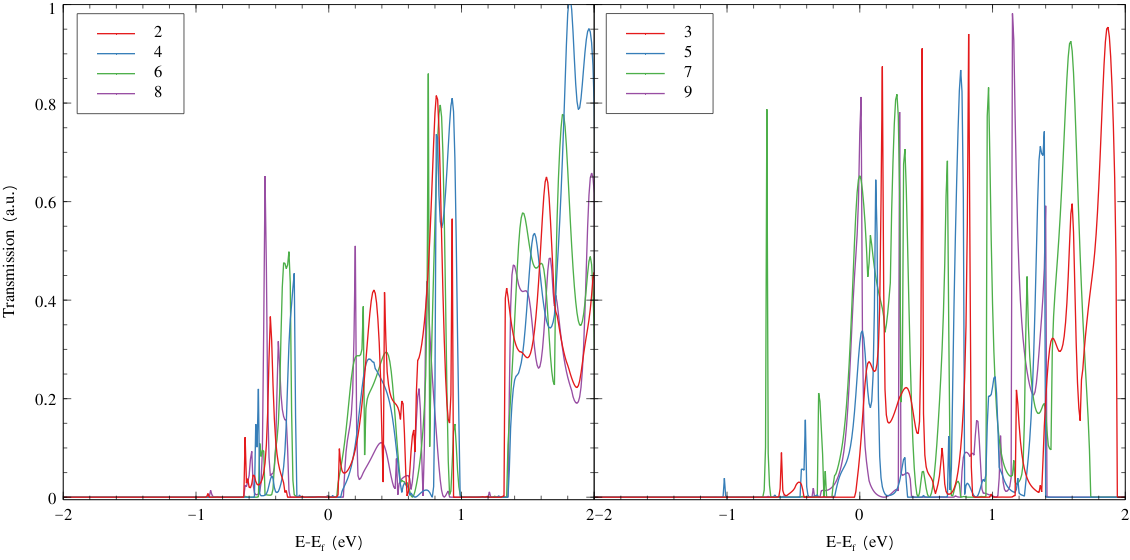}
\end{center}
\caption{Transmission functions for the carbon wire systems with NEGF-MCPDFT using the CASSCF reference wave function.  Even and odd cases are grouped together.}
\label{fig:trans_CASSCF}
\end{figure}

Transmission functions for the systems using NEGF-MCPDFT with the optimized CASSCF orbitals are given in Fig.~\ref{fig:trans_CASSCF}.  Considering first the even-length cases, the CASSCF results show some similar features to the CASCI results.  Two broad regions for transmission are evident near the Fermi level, one above and one below.  There are no peaks which overlap the Fermi level.
Like the CASCI results, the CASSCF predicts correctly these polyynic systems to be more semiconducting in character.

For the odd-length cases, the CASSCF-based results predict transmission features overlapping with the Fermi level and more transmission than CASCI, and the peaks overlap the Fermi level with an increasing amount as the chain length increases.  Again, CASSCF is predicting the correct conducting behavior for these cumulenic systems.

\begin{figure}
\begin{center}
\includegraphics[scale=.5]{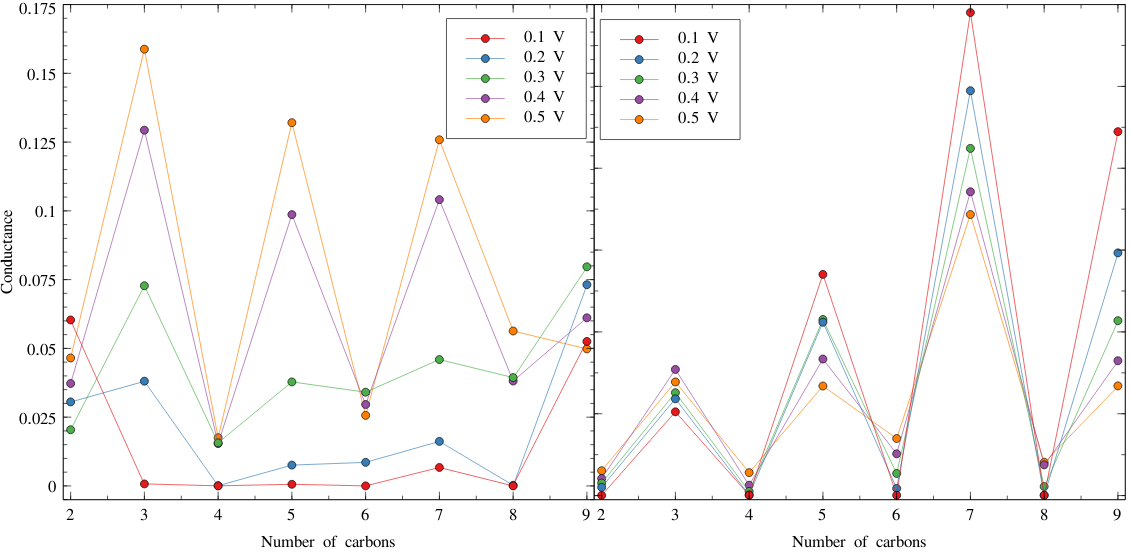}
\end{center}
\caption{Calculated conductance using NEGF-MCPDFT of carbon wire systems as a function of wire length. A: CASCI reference wave function; B: CASSCF reference wave function}
\label{fig:con_cas}
\end{figure}

Because our approach necessitates the use of an approximate Fermi level, it is important to examine the predicted conductance properties, as these results should be less susceptible to a small change in the assumed Fermi level when modest bias voltages are explored. Conductance trends for both the CASCI and CASSCF approaches were calculated for small biases (0.1-0.5~V) and are shown in Fig.~\ref{fig:con_cas}.  For both options of reference wave function, an alternating pattern emerges.  With this pattern emerging only at the higher voltages for CASCI, it suggests there may be some slight error which is introduced by the misalignment of the central and electrode regions and the use of the approximate Fermi level.  In other words, the slight shift of the transport signals to energy levels just off of the Fermi level means very small to zero bias conductances are most susceptible to shifts of the transport signals, hence the emergence of the alternating pattern only at higher voltages for the CASCI case.  The CASSCF results show the alternating pattern at all explored bias voltages.

Among the odd-length carbon cases, we see that conductance is generally not predicted to decay exponentially as the chain length increases.  This trend has also been observed experimentally. In contrast, even-length chains of more semiconducting character are experimentally observed to have exponential conductance decays.  Both the CASCI and CASSCF results do not show this, but instead predict more variable, but low, conductances as a function of length.   

Clearly, the use of a particular reference wave function and a particular active space (size and orbital character selected) can influence the predicted electron transport properties of a system.  Below we will explore more general lessons and active space design principles for systems in general, but now we want to note that despite the CASSCF active orbitals looking very poor (asymmetric, highly localized) in comparison to the delocalized and interface-containing DFT orbitals, the CASSCF results show the results most consistent with experiment.  This is perhaps a surprising result.  It appears that for this system, having a better description of the multireference correlation energy, whether it is found in the central region or in the electrode regions, is a very important factor in describing overall transport properties compared to having larger orbital densities in the interface or central regions.  Evident by the resultant orbitals and a comparison of the CASCI vs. CASSCF orbital occupation numbers, it is the electrodes (metals) that are most prone to improvement by inclusion in the active space.

\subsection{Electron transport with a DFT-like approach}

In order to further understand and contextualize our results, we performed transport calculations using a more NEGF-DFT-like approach.  In the NEGF-MCPDFT framework, a non-multiconfigurational DFT-like approach can be achieved by using a trivial active space (2e, 1o).  This reproduces NEGF-DFT results, with the caveat that it is a one-shot non-self-consistent procedure, i.e there is no optimization of the density with respect to coupling between the electrode and central regions.  Previous studies using corrected self-consistent NEGF-DFT are capable of producing the correct oscillatory behavior~\cite{prasongkit_cumulene_2010,wang_first-principles_2012,ferreira_electronic_2020,wang_exploring_2024}. However, for our purposes, it is worthwhile to explore this as we can see how the introduction of the multiconfigurational character of the CAS and PDFT approaches directly affect the transport results.

\begin{figure}
\begin{center}
\includegraphics[scale=.5]{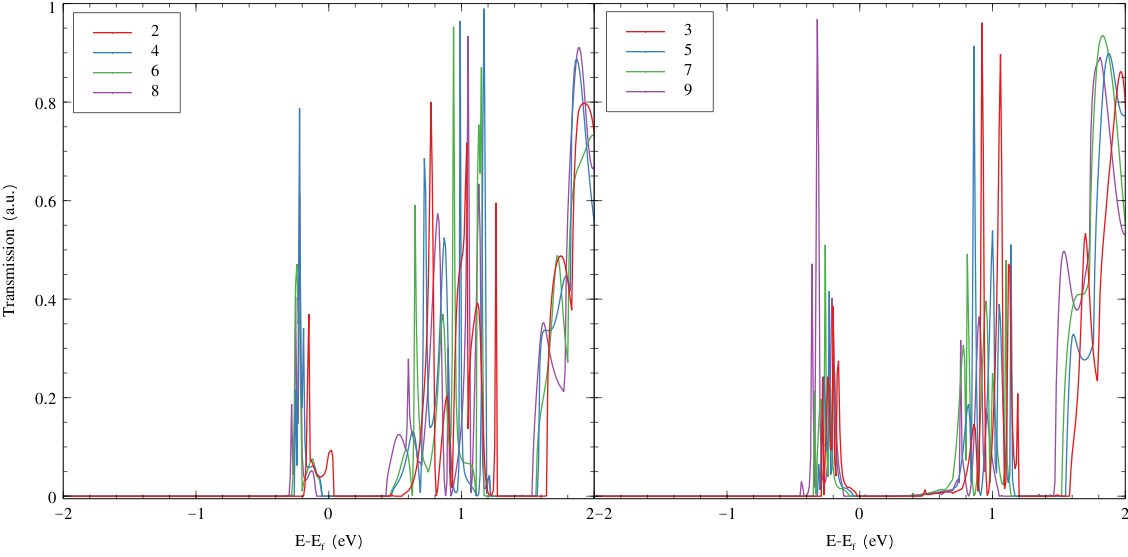}
\end{center}
\caption{Transmission of DFT... Even and odd cases are grouped together.}
\label{fig:DFT_trans}
\end{figure}

Transmission curves for the DFT approach are given in Fig.~\ref{fig:DFT_trans}.  In comparison to the CASCI and CASSCF results, some similarities and some differences are evident.  First, comparing to the CASCI case (which uses the same orbitals), we see that the even-length carbon wires have two general regions where transmission is high.  The DFT results are similar to the CASCI results, but the lower-energy transmission peak is very near (or overlapping in the three-carbon case) the Fermi level.  This, in turn, probably suggests a calculated conductance that is too high, a common error seen in non-self-consistent NEGF-DFT calculations. For the odd-length systems, larger differences are seen.  The DFT results show a greater consistency between predicted transmission as the chain length increases compared to either the CASCI or CASSCF results.  The transmission peak nearest the Fermi level, though, does not strongly overlap with it.  This suggests DFT is not predicting the higher conductance for the odd-length cumulenic structures.

A further understanding of the trends observed here can be attained by examining the orbital occupation numbers (Figs~\ref{fig:3_orb} and \ref{fig:4_orb}) in the CASCI case.  The DFT and CASCI reference wave functions use the same orbitals.  In the even-length case, we see the CASCI orbital occupations are close to the DFT occupations (only 0 or 2 are allowed in the single-determinant case).  The middle two orbitals, with occupation numbers of 1.401 and 0.598 are quite different from 2 and 0, respectively, but this is the only large difference.  Hence, the predicted even-length transmission curves for DFT and CASCI are qualitatively similar.  In contrast, the odd-length CASCI occupation numbers show greater differences from their DFT values, with four occupation numbers very near 1 (far from 0 or 2).  Hence, the transmission curves look quite different.

\begin{figure}
\begin{center}
\includegraphics[scale=.5]{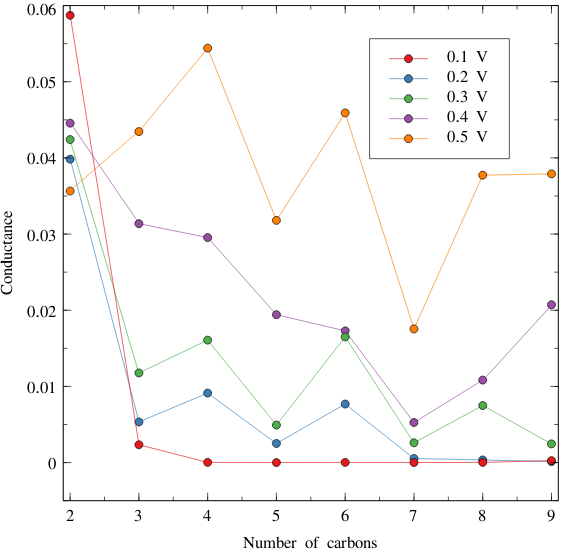}
\end{center}
\caption{Conductance of carbon wire systems using a one-shot NEGF-DFT approach.}
\label{fig:cond_dft}
\end{figure}

Like with the CASCI and CASSCF reference wave functions, we calculate the conductance as a function of chain length for the DFT case (Fig.~\ref{fig:cond_dft}.  The even-odd character alternating character is not seen in the DFT conductance results until 0.5~V, but the alternating pattern is opposite the experimental results; even-length chains show higher conductance.  The low-voltage conductance results, do, however, show a decaying pattern as a function of system length.

\subsection{Lessons on active space selection}

From these presented results, we wish to explain our insights into the definition of appropriate reference wave functions and active space designs for NEGF-MCPDFT calculations.  A key step in most multiconfigurational wave function calculations is the selection of the orbitals which will be included in the active space.  Best practices and automated tools have been developed to influence and dictate active space design, but mostly for the accurate description of static chemical properties such as reaction energetics or excited state energies. If instead one wishes to use active space wave function to describe electron transport in junctions, much less is known about what an appropriate active space looks like.  

In this study exploring one dimensional carbon systems, we find that we obtain the most reliable results when CASSCF optimized orbitals were used instead of CASCI calculations with delocalized orbitals from DFT calculations.  The CASSCF orbitals were generally more localized on the terminal electrodes, not in the central molecular region.  In one sense, this is not surprising as the electrode region, consisting of metals, exhibit more multiconfigurational character while the central region, consisting of carbons, exhibits far less multiconfigurational character.  It is surprising, however, that a good description of the electrode (provided by the CASSCF orbitals) is so important in producing transport and conductance results that are consistent with experiment.  

In previous studies involving NEGF-MCPDFT, it was suggested that essential components of a good active space for transport calculations should include orbitals which have significant character in the interface region between the electrode and central region~\cite{sand_multiconfiguration_2021}.  These studies did explore very different junctions, in particular molecules with significant multiconfigurational character (diradicals, for example).  
This current manuscript, then, clarifies our understanding of some basic active space design principles.  The active space needs to describe both the multiconfigurational character in the electrode regions and the central region (to the extent that it is present).  Allowing for orbital optimization at the CASSCF level is a good way to do this.

When exploring multiple (but related) systems (like in this manuscript, carbon wires with increasing length), the use of an active space approach presents some challenges.  Can one keep the number of active orbitals and active electrons consistent across calculations as the size of the system increases, or should the active space also increase with system size?  What results in the most fair comparison between systems?  There many not be perfect answers to these questions.  Certainly, though, constraints on the computational cost of ever increasing active spaces is a fact that might promote more uniformly-sized active spaces across systems.  This is a potential source of inconsistency from system to system, and we may see evidence of that as the CASCI odd-length transmission plot (Fig.~\ref{fig:trans_CASCI}) shows lots more deviation as the length changes when compared to DFT (Fig.~\ref{fig:DFT_trans}).  This should be kept in mind in a series of calculations such as these.

Lastly, it may be possible that there are multiple spatial regions in junction systems exhibiting significant multiconfigurational character, say both the electrode regions and the central (molecular) region.  It may be ideal or desirable to constrain an active space to particular regions in space.  One logical area for future exploration, then, is the use of previously-developed spatially-constrained active space approaches. Several groups have developed such approaches; these include the localized active space self-consistent field (LASSCF) method~\cite{hermes_multiconfigurational_2019,hermes_variational_2020,pandharkar_localized_2021,mitra_localized_2024} and fragment-based periodic approaches~\cite{lin_fragment-based_2020}. They have proved useful for embedding calculations, and they may have good applicability to transport simulations as well, particularly if there is strong correlation exhibited in the central (molecular) or interface regions.  This will be a future area of study.

\section{Conclusion}

One-dimensional carbon wire materials are an interesting class of materials which have potentially useful applications in nanoscale junction technology.  In this work, we explored the transport properties of a series of carbon wire systems with NEGF-MCPDFT, a wave-function-based non-periodic methodology which is capable of describing systems exhibiting strong multiconfigurational behavior.  In most multiconfigurational approaches, key decisions must be made, including whether to optimize the orbitals (CASSCF) or not (CASCI) and which orbitals are to be included in the active space region.  We have found that the added flexibility that is introduced by allowing for orbital optimization can have a strong influence on the ability of the NEGF-MCPDFT method to predict results consistent with experimental observable transport properties, despite the CASSCF orbitals looking visually poor.  We attribute this improvement to an enhanced description of the highly multiconfigurational electrode regions, consisting of metal atoms.  In the future, a fragmented active space approach may be necessary to adequately describe systems which exhibit strong multiconfigurational character in both electrode and central molecular regions, and the applicability and feasibility of such approaches will be studied in future work. 

%Strange result that the bad looking orbitals get the best results.

%\section{Associated Content}
%\subsection{Supporting Information}

%\section{Supplementary Material}

\section{Acknowledgment}
A.M.S acknowledges support from the National Science Foundation grant 2154833.  E.P.H. acknowledges support from the National Science Foundation grant 2154832.

\bibliography{ref}

@string{S = "Science"}

@article{al-backri_electronic_2014,
	title = {Electronic properties of linear carbon chains: {Resolving} the controversy},
	volume = {140},
	issn = {0021-9606},
	shorttitle = {Electronic properties of linear carbon chains},
	url = {https://doi.org/10.1063/1.4867635},
	doi = {10.1063/1.4867635},
	number = {10},
	urldate = {2026-07-24},
	journal = {J. Chem. Phys.},
	author = {Al-Backri, Amaal and Zólyomi, Viktor and Lambert, Colin J.},
	month = mar,
	year = {2014},
	pages = {104306},
}

@article{kumar_effect_2015,
	author  = {Kumar, Maneesh and Husain, Mudassir M.},
	title   = {Effect of {Electrodes} on {Negative} {Differential} {Resistance} in {Carbon} {Nanowire}: {A} {First}-{Principles} {Study}},
	journal = {Advanced Science Letters},
	volume  = {21},
	number  = {9},
	pages   = {2768--2771},
	month   = sep,
	year    = {2015},
	doi     = {10.1166/asl.2015.6343}
}

@article{wang_first-principles_2012,
	author    = {Wang, Bin and Wei, Yadong and Wang, Jian},
	title     = {First-principles calculation of the {Andreev} conductance of carbon wires},
	journal   = {Phys. Rev. B},
	publisher = {American Physical Society},
	volume    = {86},
	number    = {3},
	pages     = {035414},
	month     = jul,
	year      = {2012},
	doi       = {10.1103/PhysRevB.86.035414},
	url       = {https://link.aps.org/doi/10.1103/PhysRevB.86.035414},
	urldate   = {2026-08-23}
}

@article{prasongkit_cumulene_2010,
	author    = {Prasongkit, J. and Grigoriev, A. and Wendin, G. and Ahuja, Rajeev},
	title     = {Cumulene molecular wire conductance from first principles},
	journal   = {Phys. Rev. B},
	publisher = {American Physical Society},
	volume    = {81},
	number    = {11},
	pages     = {115404},
	month     = mar,
	year      = {2010},
	doi       = {10.1103/PhysRevB.81.115404},
	url       = {https://link.aps.org/doi/10.1103/PhysRevB.81.115404},
	urldate   = {2026-08-23}
}

@article{mu_ab_2023,
	author  = {Mu, Yi and Yu, Jie and Hu, Rui and Wang, Cui-Hong and Cheng, Cai and Hou, Bang-Pin},
	title   = {Ab initio study revealing remarkable oscillatory effects and negative differential resistance in the molecular device of silicon carbide chains},
	journal = {Phys. Chem. Chem. Phys.},
	volume  = {25},
	number  = {19},
	pages   = {13265--13274},
	month   = may,
	year    = {2023},
	issn    = {1463-9076},
	doi     = {10.1039/d2cp05677a},
	url     = {https://doi.org/10.1039/d2cp05677a},
	urldate = {2026-08-23}
}

@article{balakrishnan_polyyne-metal_2021,
	author  = {Balakrishnan, AbhayRam and Shankar, R. and Vijayakumar, S.},
	title   = {Polyyne-metal complexes for use in molecular wire applications: {A} {DFT} insight},
	journal = {Computational and Theoretical Chemistry},
	volume  = {1202},
	pages   = {113328},
	month   = aug,
	year    = {2021},
	issn    = {2210-271X},
	doi     = {10.1016/j.comptc.2021.113328},
	url     = {https://www.sciencedirect.com/science/article/pii/S2210271X21001869},
	urldate = {2026-08-23}
}

@article{zang_cumulene_2020,
	author  = {Zang, Yaping and Fu, Tianren and Zou, Qi and Ng, Fay and Li, Hexing and Steigerwald, Michael L. and Nuckolls, Colin and Venkataraman, Latha},
	title   = {Cumulene {Wires} {Display} {Increasing} {Conductance} with {Increasing} {Length}},
	journal = {Nano Lett.},
	volume  = {20},
	number  = {11},
	pages   = {8415--8419},
	month   = oct,
	year    = {2020},
	issn    = {1530-6984},
	doi     = {10.1021/acs.nanolett.0c03794},
	url     = {https://doi.org/10.1021/acs.nanolett.0c03794},
	urldate = {2026-08-23}
}

@article{balakrishnan_dft_2020,
	author    = {Balakrishnan, AbhayRam and Shankar, R. and Vijayakumar, S.},
	title     = {{DFT} approach on stability and conductance of nine different polyyne and cumulene molecules},
	journal   = {Molecular Physics},
	publisher = {Taylor \& Francis},
	volume    = {118},
	number    = {2},
	pages     = {e1601785},
	month     = jan,
	year      = {2020},
	issn      = {0026-8976},
	doi       = {10.1080/00268976.2019.1601785},
	url       = {https://doi.org/10.1080/00268976.2019.1601785},
	urldate   = {2026-08-23}
}

@article{garner_three_2020,
	author  = {Garner, Marc H. and Bro-Jørgensen, William and Solomon, Gemma C.},
	title   = {Three {Distinct} {Torsion} {Profiles} of {Electronic} {Transmission} through {Linear} {Carbon} {Wires}},
	journal = {J. Phys. Chem. C},
	volume  = {124},
	number  = {35},
	pages   = {18968--18982},
	month   = aug,
	year    = {2020},
	issn    = {1932-7447},
	doi     = {10.1021/acs.jpcc.0c07051},
	url     = {https://doi.org/10.1021/acs.jpcc.0c07051},
	urldate = {2026-08-23}
}

@article{xu_unusual_2019,
	author   = {Xu, Wenjun and Leary, Edmund and Hou, Songjun and Sangtarash, Sara and González, M. Teresa and Rubio-Bollinger, Gabino and Wu, Qingqing and Sadeghi, Hatef and Tejerina, Lara and Christensen, Kirsten E. and Agraït, Nicolás and Higgins, Simon J. and Lambert, Colin J. and Nichols, Richard J. and Anderson, Harry L.},
	title    = {Unusual {Length} {Dependence} of the {Conductance} in {Cumulene} {Molecular} {Wires}},
	journal  = {Angewandte Chemie},
	volume   = {131},
	number   = {25},
	pages    = {8466--8470},
	year     = {2019},
	language = {en},
	issn     = {1521-3757},
	doi      = {10.1002/ange.201901228},
	url      = {https://onlinelibrary.wiley.com/doi/abs/10.1002/ange.201901228},
	urldate  = {2026-08-23}
}

@article{sarbadhikary_magnetic_2018,
	author  = {Sarbadhikary, Prodipta and Shil, Suranjan and Misra, Anirban},
	title   = {Magnetic and transport properties of conjugated and cumulated molecules: the {$\pi$}-system enlightens part of the story},
	journal = {Phys. Chem. Chem. Phys.},
	volume  = {20},
	number  = {14},
	pages   = {9364--9375},
	month   = apr,
	year    = {2018},
	issn    = {1463-9076},
	doi     = {10.1039/c7cp06113g},
	url     = {https://doi.org/10.1039/c7cp06113g},
	urldate = {2026-08-23}
}

@article{garner_reverse_2018,
	author  = {Garner, Marc H. and Bro-Jørgensen, William and Pedersen, Pernille D. and Solomon, Gemma C.},
	title   = {Reverse {Bond}-{Length} {Alternation} in {Cumulenes}: {Candidates} for {Increasing} {Electronic} {Transmission} with {Length}},
	journal = {J. Phys. Chem. C},
	volume  = {122},
	number  = {47},
	pages   = {26777--26789},
	month   = aug,
	year    = {2018},
	issn    = {1932-7447},
	doi     = {10.1021/acs.jpcc.8b05661},
	url     = {https://doi.org/10.1021/acs.jpcc.8b05661},
	urldate = {2026-08-23}
}

@article{fang_electronic_2011,
	author  = {Fang, Changfeng and Cui, Bin and Xu, Yuqing and Ji, Guomin and Liu, Desheng and Xie, Shijie},
	title   = {Electronic transport properties of carbon chains between {Au} and {Ag} electrodes: {A} first-principles study},
	journal = {Physics Letters A},
	volume  = {375},
	number  = {41},
	pages   = {3618--3623},
	month   = sep,
	year    = {2011},
	issn    = {0375-9601},
	doi     = {10.1016/j.physleta.2011.08.032},
	url     = {https://www.sciencedirect.com/science/article/pii/S037596011101005X},
	urldate = {2026-08-23}
}

@article{shi_confined_2016,
	author    = {Shi, Lei and Rohringer, Philip and Suenaga, Kazu and Niimi, Yoshiko and Kotakoski, Jani and Meyer, Jannik C. and Peterlik, Herwig and Wanko, Marius and Cahangirov, Seymur and Rubio, Angel and Lapin, Zachary J. and Novotny, Lukas and Ayala, Paola and Pichler, Thomas},
	title     = {Confined linear carbon chains as a route to bulk carbyne},
	journal   = {Nature Mater.},
	publisher = {Nature Publishing Group},
	volume    = {15},
	number    = {6},
	pages     = {634--639},
	month     = jun,
	year      = {2016},
	issn      = {1476-4660},
	doi       = {10.1038/nmat4617},
	url       = {https://www.nature.com/articles/nmat4617},
	urldate   = {2026-07-24}
}

@article{morris_charge_2026,
	author    = {Morris, James M. F. and Potter, Jarred and Gorenskaia, Elena and Abram, R. Tom and Naher, Masnun and Spano, Chiara E. and Listo, Roberto and Dixon, Eloise L. and Sil, Amit and Rousset, \'{E}lodie and Higgins, Simon J. and Nichols, Richard J. and Low, Paul J. and Vezzoli, Andrea},
	title     = {Charge transport through linear carbon atomic chains},
	journal   = {Nat. Chem.},
	publisher = {Nature Publishing Group},
	volume    = {18},
	pages     = {1--8},
	month     = jun,
	year      = {2026},
	issn      = {1755-4349},
	doi       = {10.1038/s41557-026-02175-w},
	url       = {https://www.nature.com/articles/s41557-026-02175-w},
	urldate   = {2026-07-24}
}

@article{wang_oligoyne_2009,
	author   = {Wang, Changsheng and Batsanov, Andrei S. and Bryce, Martin R. and Martín, Santiago and Nichols, Richard J. and Higgins, Simon J. and García-Suárez, Víctor M. and Lambert, Colin J.},
	title    = {Oligoyne {Single} {Molecule} {Wires}},
	journal  = {J. Am. Chem. Soc.},
	volume   = {131},
	number   = {43},
	pages    = {15647--15654},
	month    = oct,
	year     = {2009},
	issn     = {0002-7863},
	doi      = {10.1021/ja9061129},
	url      = {https://doi.org/10.1021/ja9061129},
	urldate  = {2026-08-12}
}

@article{gao_loss_2020,
	author    = {Gao, Yueze and Hou, Yuxuan and Gordillo Gámez, Fernando and Ferguson, Mike J. and Casado, Juan and Tykwinski, Rik R.},
	title     = {The loss of endgroup effects in long pyridyl-endcapped oligoynes on the way to carbyne},
	journal   = {Nat. Chem.},
	publisher = {Nature Publishing Group},
	volume    = {12},
	number    = {12},
	pages     = {1143--1149},
	month     = dec,
	year      = {2020},
	issn      = {1755-4349},
	doi       = {10.1038/s41557-020-0550-0},
	url       = {https://www.nature.com/articles/s41557-020-0550-0},
	urldate   = {2026-08-23}
}

@article{kaiser_sp-hybridized_2019,
	author    = {Kaiser, Katharina and Scriven, Lorel M. and Schulz, Fabian and Gawel, Przemyslaw and Gross, Leo and Anderson, Harry L.},
	title     = {An {$sp$}-hybridized molecular carbon allotrope, cyclo[18]carbon},
	journal   = {Science},
	publisher = {American Association for the Advancement of Science},
	volume    = {365},
	number    = {6459},
	pages     = {1299--1301},
	month     = sep,
	year      = {2019},
	doi       = {10.1126/science.aay1914},
	url       = {https://www.science.org/doi/10.1126/science.aay1914},
	urldate   = {2026-08-23}
}

@article{chalifoux_synthesis_2008,
	author     = {Chalifoux, Wesley A. and Tykwinski, Rik R.},
	title      = {Synthesis of extended polyynes: {Toward} carbyne},
	journal    = {C. R. Chim.},
	volume     = {12},
	number     = {3-4},
	pages      = {341--358},
	month      = dec,
	year       = {2008},
	issn       = {1878-1543},
	doi        = {10.1016/j.crci.2008.10.004},
	url        = {https://comptes-rendus.academie-sciences.fr/chimie/articles/10.1016/j.crci.2008.10.004/},
	urldate    = {2026-08-23}
}

@article{szafert_update_2006,
	author     = {Szafert, Slawomir and Gladysz, J. A.},
	title      = {Update 1 of: {Carbon} in {One} {Dimension}: {Structural} {Analysis} of the {Higher} {Conjugated} {Polyynes}},
	journal    = {Chem. Rev.},
	volume     = {106},
	number     = {11},
	pages      = {PR1--PR33},
	month      = nov,
	year       = {2006},
	issn       = {0009-2665},
	doi        = {10.1021/cr068016g},
	url        = {https://doi.org/10.1021/cr068016g},
	urldate    = {2026-08-23}
}

@article{eisler_polyynes_2005,
	author   = {Eisler, Sara and Slepkov, Aaron D. and Elliott, Erin and Luu, Thanh and McDonald, Robert and Hegmann, Frank A. and Tykwinski, Rik R.},
	title    = {Polyynes as a {Model} for {Carbyne}: {Synthesis}, {Physical} {Properties}, and {Nonlinear} {Optical} {Response}},
	journal  = {J. Am. Chem. Soc.},
	volume   = {127},
	number   = {8},
	pages    = {2666--2676},
	month    = feb,
	year     = {2005},
	issn     = {0002-7863},
	doi      = {10.1021/ja044526l},
	url      = {https://doi.org/10.1021/ja044526l},
	urldate  = {2026-08-23}
}

@article{schermann_dicyanopolyynes_1997,
	author   = {Schermann, Günther and Grösser, Thomas and Hampel, Frank and Hirsch, Andreas},
	title    = {Dicyanopolyynes: {A} {Homologous} {Series} of {End}-{Capped} {Linear} {$sp$} {Carbon}},
	journal  = {Chem. Eur. J.},
	volume   = {3},
	number   = {7},
	pages    = {1105--1112},
	year     = {1997},
	issn     = {1521-3765},
	doi      = {10.1002/chem.19970030718},
	url      = {https://onlinelibrary.wiley.com/doi/abs/10.1002/chem.19970030718},
	urldate  = {2026-08-23}
}

@article{wang_exploring_2024,
	author    = {Wang, Lijun and Zhou, Liping and Wang, Xuefeng and You, Wenlong},
	title     = {Exploring the {Odd}–{Even} {Effect}, {Current} {Stabilization}, and {Negative} {Differential} {Resistance} in {Carbon}-{Chain}-{Based} {Molecular} {Devices}},
	journal   = {Electronics},
	publisher = {MDPI},
	volume    = {13},
	number    = {9},
	pages     = {1764},
	month     = jan,
	year      = {2024},
	issn      = {2079-9292},
	doi       = {10.3390/electronics13091764},
	url       = {https://www.mdpi.com/2079-9292/13/9/1764},
	urldate   = {2025-05-15}
}

@article{ferreira_electronic_2020,
	author   = {Ferreira, D. F. S. and Moura-Moreira, M. and Corrêa, S. M. and {da Silva Jr}, C. A. B. and {Del Nero}, J.},
	title    = {Electronic transport in {1D} system with coupling atomic-size nickel electrodes and carbon wires},
	journal  = {Mater. Sci. Eng. B},
	volume   = {262},
	pages    = {114681},
	month    = dec,
	year     = {2020},
	issn     = {0921-5107},
	doi      = {10.1016/j.mseb.2020.114681},
	url      = {https://www.sciencedirect.com/science/article/pii/S0921510720301884},
	urldate  = {2025-05-19}
}

@article{ren_collection_2024,
	author   = {Ren, Jinlong and Li, Tianchen and Li, Zhuang and Kong, Decheng and Shan, Guangcun and Dou, KunPeng},
	title    = {Collection of unconventional transport phenomena: natural obstacle or vibrant guiding principle for the design of molecular junctions?},
	journal  = {AAPPS Bull.},
	volume   = {34},
	number   = {1},
	pages    = {3},
	month    = jan,
	year     = {2024},
	issn     = {2309-4710},
	doi      = {10.1007/s43673-023-00110-6},
	url      = {https://doi.org/10.1007/s43673-023-00110-6},
	urldate  = {2025-05-21}
}

@article{casari_carbyne_2018,
	author  = {Casari, C. S. and Milani, A.},
	title   = {Carbyne: from the elusive allotrope to stable carbon atom wires},
	journal = {MRS Commun.},
	volume  = {8},
	number  = {2},
	pages   = {207--219},
	month   = jun,
	year    = {2018},
	issn    = {2159-6867},
	doi     = {10.1557/mrc.2018.48},
	url     = {https://doi.org/10.1557/mrc.2018.48},
	urldate = {2026-08-03}
}

@article{hirsch_era_2010,
	author    = {Hirsch, Andreas},
	title     = {The era of carbon allotropes},
	journal   = {Nature Mater.},
	publisher = {Nature Publishing Group},
	volume    = {9},
	number    = {11},
	pages     = {868--871},
	month     = nov,
	year      = {2010},
	issn      = {1476-4660},
	doi       = {10.1038/nmat2885},
	url       = {https://www.nature.com/articles/nmat2885},
	urldate   = {2026-08-12}
}

@article{bryce_review_2021,
	author  = {Bryce, Martin R.},
	title   = {A review of functional linear carbon chains (oligoynes, polyynes, cumulenes) and their applications as molecular wires in molecular electronics and optoelectronics},
	journal = {J. Mater. Chem. C},
	volume  = {9},
	number  = {33},
	pages   = {10524--10546},
	month   = sep,
	year    = {2021},
	issn    = {2050-7526},
	doi     = {10.1039/d1tc01406d},
	url     = {https://doi.org/10.1039/d1tc01406d},
	urldate = {2026-08-12}
}

@article{casari_carbon-atom_2016,
	author  = {Casari, C. S. and Tommasini, M. and Tykwinski, R. R. and Milani, A.},
	title   = {Carbon-atom wires: 1-{D} systems with tunable properties},
	journal = {Nanoscale},
	volume  = {8},
	number  = {8},
	pages   = {4414--4435},
	month   = feb,
	year    = {2016},
	issn    = {2040-3364},
	doi     = {10.1039/c5nr06175j},
	url     = {https://doi.org/10.1039/c5nr06175j},
	urldate = {2026-08-12}
}

@article{milani_carbon_2006,
	author    = {Milani, Alberto and Tommasini, Matteo and Del Zoppo, Mirella and Castiglioni, Chiara and Zerbi, Giuseppe},
	title     = {Carbon nanowires: {Phonon} and {$\pi$}-electron confinement},
	journal   = {Phys. Rev. B},
	publisher = {American Physical Society},
	volume    = {74},
	number    = {15},
	pages     = {153418},
	month     = oct,
	year      = {2006},
	doi       = {10.1103/PhysRevB.74.153418},
	url       = {https://link.aps.org/doi/10.1103/PhysRevB.74.153418},
	urldate   = {2026-08-23}
}

@article{bergner_ab_1993,
	author    = {Bergner, Andreas and Dolg, Michael and Küchle, Wolfgang and Stoll, Hermann and Preuß, Heinzwerner},
	title     = {Ab initio energy-adjusted pseudopotentials for elements of groups 13--17},
	journal   = {Mol. Phys.},
	publisher = {Taylor \& Francis},
	volume    = {80},
	number    = {6},
	pages     = {1431--1441},
	month     = dec,
	year      = {1993},
	issn      = {0026-8976},
	doi       = {10.1080/00268979300103121},
	url       = {https://doi.org/10.1080/00268979300103121},
	urldate   = {2026-08-23}
}

@article{soler_siesta_2002,
	author    = {Soler, José M. and Artacho, Emilio and Gale, Julian D. and García, Alberto and Junquera, Javier and Ordejón, Pablo and Sánchez-Portal, Daniel},
	title     = {The {SIESTA} method for ab initio order-{N} materials simulation},
	journal   = {J. Phys. Condens. Matter},
	volume    = {14},
	number    = {11},
	pages     = {2745--2779},
	month     = mar,
	year      = {2002},
	issn      = {0953-8984},
	doi       = {10.1088/0953-8984/14/11/302},
	url       = {https://doi.org/10.1088/0953-8984/14/11/302},
	urldate   = {2026-08-23}
}

@article{li_manni_openmolcas_2023,
	title = {The {OpenMolcas} {Web}: {A} {Community}-{Driven} {Approach} to {Advancing} {Computational} {Chemistry}},
	volume = {19},
	issn = {1549-9618},
	shorttitle = {The {OpenMolcas} {Web}},
	url = {https://doi.org/10.1021/acs.jctc.3c00182},
	doi = {10.1021/acs.jctc.3c00182},
	number = {20},
	urldate = {2026-08-23},
	journal = {J. Chem. Theory Comput.},
	author = {Li Manni, Giovanni and Fdez. Galván, Ignacio and Alavi, Ali and Aleotti, Flavia and Aquilante, Francesco and Autschbach, Jochen and Avagliano, Davide and Baiardi, Alberto and Bao, Jie J. and Battaglia, Stefano and Birnoschi, Letitia and Blanco-González, Alejandro and Bokarev, Sergey I. and Broer, Ria and Cacciari, Roberto and Calio, Paul B. and Carlson, Rebecca K. and Carvalho Couto, Rafael and Cerdán, Luis and Chibotaru, Liviu F. and Chilton, Nicholas F. and Church, Jonathan Richard and Conti, Irene and Coriani, Sonia and Cuéllar-Zuquin, Juliana and Daoud, Razan E. and Dattani, Nike and Decleva, Piero and de Graaf, Coen and Delcey, Mickaël G. and De Vico, Luca and Dobrautz, Werner and Dong, Sijia S. and Feng, Rulin and Ferré, Nicolas and Filatov(Gulak), Michael and Gagliardi, Laura and Garavelli, Marco and González, Leticia and Guan, Yafu and Guo, Meiyuan and Hennefarth, Matthew R. and Hermes, Matthew R. and Hoyer, Chad E. and Huix-Rotllant, Miquel and Jaiswal, Vishal Kumar and Kaiser, Andy and Kaliakin, Danil S. and Khamesian, Marjan and King, Daniel S. and Kochetov, Vladislav and Krośnicki, Marek and Kumaar, Arpit Arun and Larsson, Ernst D. and Lehtola, Susi and Lepetit, Marie-Bernadette and Lischka, Hans and López Ríos, Pablo and Lundberg, Marcus and Ma, Dongxia and Mai, Sebastian and Marquetand, Philipp and Merritt, Isabella C. D. and Montorsi, Francesco and Mörchen, Maximilian and Nenov, Artur and Nguyen, Vu Ha Anh and Nishimoto, Yoshio and Oakley, Meagan S. and Olivucci, Massimo and Oppel, Markus and Padula, Daniele and Pandharkar, Riddhish and Phung, Quan Manh and Plasser, Felix and Raggi, Gerardo and Rebolini, Elisa and Reiher, Markus and Rivalta, Ivan and Roca-Sanjuán, Daniel and Romig, Thies and Safari, Arta Anushirwan and Sánchez-Mansilla, Aitor and Sand, Andrew M. and Schapiro, Igor and Scott, Thais R. and Segarra-Martí, Javier and Segatta, Francesco and Sergentu, Dumitru-Claudiu and Sharma, Prachi and Shepard, Ron and Shu, Yinan and Staab, Jakob K. and Straatsma, Tjerk P. and Sørensen, Lasse Kragh and Tenorio, Bruno Nunes Cabral and Truhlar, Donald G. and Ungur, Liviu and Vacher, Morgane and Veryazov, Valera and Voß, Torben Arne and Weser, Oskar and Wu, Dihua and Yang, Xuchun and Yarkony, David and Zhou, Chen and Zobel, J. Patrick and Lindh, Roland},
	month = may,
	year = {2023},
	pages = {6933--6991},
}

@article{mitra_localized_2024,
	title = {The {Localized} {Active} {Space} {Method} with {Unitary} {Selective} {Coupled} {Cluster}},
	volume = {20},
	issn = {1549-9618},
	url = {https://doi.org/10.1021/acs.jctc.4c00528},
	doi = {10.1021/acs.jctc.4c00528},
	number = {18},
	urldate = {2026-08-23},
	journal = {J. Chem. Theory Comput.},
	author = {Mitra, Abhishek and D’Cunha, Ruhee and Wang, Qiaohong and Hermes, Matthew R. and Alexeev, Yuri and Gray, Stephen K. and Otten, Matthew and Gagliardi, Laura},
	month = sep,
	year = {2024},
	pages = {7865--7875},
}

@article{lin_fragment-based_2020,
	title = {Fragment-{Based} {Restricted} {Active} {Space} {Configuration} {Interaction} with {Second}-{Order} {Corrections} {Embedded} in {Periodic} {Hartree}–{Fock} {Wave} {Function}},
	volume = {16},
	issn = {1549-9618},
	url = {https://doi.org/10.1021/acs.jctc.0c00576},
	doi = {10.1021/acs.jctc.0c00576},
	number = {11},
	urldate = {2026-08-23},
	journal = {J. Chem. Theory Comput.},
	author = {Lin, Hung-Hsuan and Maschio, Lorenzo and Kats, Daniel and Usvyat, Denis and Heine, Thomas},
	month = oct,
	year = {2020},
	pages = {7100--7108},
}

@article{cossaboon_assessing_2024,
	title = {Assessing the importance of multireference correlation in predicting reversed conductance decay},
	volume = {26},
	issn = {1463-9084},
	url = {https://pubs.rsc.org/en/content/articlelanding/2024/cp/d3cp01110k},
	doi = {10.1039/D3CP01110K},
	language = {en},
	number = {8},
	urldate = {2025-03-21},
	journal = {Phys. Chem. Chem. Phys.},
	publisher = {The Royal Society of Chemistry},
	author = {Cossaboon, Tanner A. and Kazmi, Samir and Tineo, Matthew and Hoy, Erik P.},
	month = feb,
	year = {2024},
	pages = {6696--6707},
}

@article{jr_gaussian_1989,
	title = {Gaussian basis sets for use in correlated molecular calculations. {I}. {The} atoms boron through neon and hydrogen},
	volume = {90},
	issn = {0021-9606, 1089-7690},
	url = {http://scitation.aip.org/content/aip/journal/jcp/90/2/10.1063/1.456153},
	doi = {10.1063/1.456153},
	number = {2},
	urldate = {2016-07-06},
	journal = {The Journal of Chemical Physics},
	author = {Jr, Thom H. Dunning},
	month = jan,
	year = {1989},
	pages = {1007--1023},
}

@incollection{dunning_gaussian_1977,
	address = {Boston, MA},
	title = {Gaussian {Basis} {Sets} for {Molecular} {Calculations}},
	isbn = {978-1-4757-0887-5},
	url = {https://doi.org/10.1007/978-1-4757-0887-5_1},
	doi = {10.1007/978-1-4757-0887-5_1},
	booktitle = {Methods of {Electronic} {Structure} {Theory}},
	publisher = {Springer US},
	author = {Dunning, Thom. H. and Hay, P. Jeffrey},
	editor = {Schaefer, Henry F.},
	year = {1977},
	pages = {1--27},
}

@article{gandus_strongly_2026,
	title = {Strongly correlated physics in organic open-shell quantum systems},
	volume = {8},
	url = {https://link.aps.org/doi/10.1103/hlbf-5llp},
	doi = {10.1103/hlbf-5llp},
	number = {2},
	urldate = {2026-06-02},
	journal = {Phys. Rev. Res.},
	publisher = {American Physical Society},
	author = {Gandus, G. and Jayaraj, A. and Passerone, D. and Stadler, R. and Luisier, M. and Valli, A.},
	month = may,
	year = {2026},
	pages = {023221},
}

@article{yang_linear_2007,
	title = {Linear {Cn} {Clusters}: {Are} {They} {Acetylenic} or {Cumulenic}?},
	volume = {112},
	issn = {1089-5639},
	shorttitle = {Linear {Cn} {Clusters}},
	url = {https://doi.org/10.1021/jp076805b},
	doi = {10.1021/jp076805b},
	number = {1},
	urldate = {2026-08-03},
	journal = {J. Phys. Chem. A},
	author = {Yang, Shujiang and Kertesz, Miklos},
	month = dec,
	year = {2007},
	pages = {146--151},
}

@article{kovacevic_luscus_2015,
	title = {Luscus: molecular viewer and editor for {MOLCAS}},
	volume = {7},
	issn = {1758-2946},
	shorttitle = {Luscus},
	url = {https://doi.org/10.1186/s13321-015-0060-z},
	doi = {10.1186/s13321-015-0060-z},
	language = {en},
	number = {1},
	urldate = {2026-08-07},
	journal = {J Cheminform},
	author = {Kova\v{c}evi\'{c}, Goran and Veryazov, Valera},
	month = apr,
	year = {2015},
	pages = {16},
}

@article{hjorth_larsen_atomic_2017,
	title = {The atomic simulation environment—a {Python} library for working with atoms},
	volume = {29},
	issn = {0953-8984},
	url = {https://doi.org/10.1088/1361-648X/aa680e},
	doi = {10.1088/1361-648X/aa680e},
	language = {en},
	number = {27},
	urldate = {2026-08-07},
	journal = {J. Phys.: Condens. Matter},
	publisher = {IOP Publishing},
	author = {Hjorth Larsen, Ask and Jørgen Mortensen, Jens and Blomqvist, Jakob and Castelli, Ivano E and Christensen, Rune and Dułak, Marcin and Friis, Jesper and Groves, Michael N and Hammer, Bj\ork and Hargus, Cory and Hermes, Eric D and Jennings, Paul C and Bjerre Jensen, Peter and Kermode, James and Kitchin, John R and Leonhard Kolsbjerg, Esben and Kubal, Joseph and Kaasbjerg, Kristen and Lysgaard, Steen and Bergmann Maronsson, J\'{o}n and Maxson, Tristan and Olsen, Thomas and Pastewka, Lars and Peterson, Andrew and Rostgaard, Carsten and Schiøtz, Jakob and Sch\"{u}tt, Ole and Strange, Mikkel and Thygesen, Kristian S and Vegge, Tejs and Vilhelmsen, Lasse and Walter, Michael and Zeng, Zhenhua and Jacobsen, Karsten W},
	month = jun,
	year = {2017},
	pages = {273002},
}

@article{pulay_uhf_1988,
	title = {{UHF} natural orbitals for defining and starting {MC}‐{SCF} calculations},
	volume = {88},
	issn = {0021-9606},
	url = {https://doi.org/10.1063/1.454704},
	doi = {10.1063/1.454704},
	number = {8},
	urldate = {2026-08-07},
	journal = {J. Chem. Phys.},
	author = {Pulay, Peter and Hamilton, Tracy P.},
	month = apr,
	year = {1988},
	pages = {4926--4933},
}

@article{veryazov_how_2011,
	title = {How to select active space for multiconfigurational quantum chemistry?},
	volume = {111},
	copyright = {Copyright  2011 Wiley Periodicals, Inc.},
	issn = {1097-461X},
	url = {https://onlinelibrary.wiley.com/doi/abs/10.1002/qua.23068},
	doi = {10.1002/qua.23068},
	language = {en},
	number = {13},
	urldate = {2026-08-07},
	journal = {International Journal of Quantum Chemistry},
	author = {Veryazov, Valera and Malmqvist, Per {\AA}ke and Roos, Bj\"{o}rn O.},
	year = {2011},
	pages = {3329--3338},
}

@article{stein_automated_2016,
	title = {Automated {Selection} of {Active} {Orbital} {Spaces}},
	volume = {12},
	issn = {1549-9618},
	url = {https://dx.doi.org/10.1021/acs.jctc.6b00156},
	doi = {10.1021/acs.jctc.6b00156},
	language = {en},
	number = {4},
	urldate = {2026-08-07},
	journal = {J. Chem. Theory Comput.},
	publisher = {American Chemical Society},
	author = {Stein, Christopher J. and Reiher, Markus},
	month = apr,
	year = {2016},
	pages = {1760--1771},
}

@article{bao_automatic_2018,
	title = {Automatic {Selection} of an {Active} {Space} for {Calculating} {Electronic} {Excitation} {Spectra} by {MS}-{CASPT2} or {MC}-{PDFT}},
	volume = {14},
	issn = {1549-9618},
	url = {https://dx.doi.org/10.1021/acs.jctc.8b00032},
	doi = {10.1021/acs.jctc.8b00032},
	language = {en},
	number = {4},
	urldate = {2026-08-07},
	journal = {J. Chem. Theory Comput.},
	publisher = {American Chemical Society},
	author = {Bao, Jie J. and Dong, Sijia S. and Gagliardi, Laura and Truhlar, Donald G.},
	month = apr,
	year = {2018},
	pages = {2017--2025},
}

@article{sayfutyarova_constructing_2019,
	title = {Constructing {Molecular} {$\pi$}‑{Orbital} {Active} {Spaces} for {Multireference} {Calculations} of {Conjugated} {Systems}},
	volume = {15},
	issn = {1549-9618},
	url = {https://dx.doi.org/10.1021/acs.jctc.8b01196},
	doi = {10.1021/acs.jctc.8b01196},
	language = {en},
	number = {3},
	urldate = {2026-08-07},
	journal = {J. Chem. Theory Comput.},
	publisher = {American Chemical Society},
	author = {Sayfutyarova, Elvira R. and Hammes-Schiffer, Sharon},
	month = mar,
	year = {2019},
	pages = {1679--1689},
}

@article{khedkar_active_2019,
	title = {Active {Space} {Selection} {Based} on {Natural} {Orbital} {Occupation} {Numbers} from n‑{Electron} {Valence} {Perturbation} {Theory}},
	volume = {15},
	issn = {1549-9618},
	url = {https://dx.doi.org/10.1021/acs.jctc.8b01293},
	doi = {10.1021/acs.jctc.8b01293},
	language = {en},
	number = {6},
	urldate = {2026-08-07},
	journal = {J. Chem. Theory Comput.},
	publisher = {American Chemical Society},
	author = {Khedkar, Abhishek and Roemelt, Michael},
	month = jun,
	year = {2019},
	pages = {3522--3536},
}

@article{kolodzeiski_automated_2023,
	title = {Automated, {Consistent}, and {Even}-{Handed} {Selection} of {Active} {Orbital} {Spaces} for {Quantum} {Embedding}},
	volume = {19},
	issn = {1549-9618},
	url = {https://doi.org/10.1021/acs.jctc.3c00653},
	doi = {10.1021/acs.jctc.3c00653},
	number = {19},
	urldate = {2026-08-07},
	journal = {J. Chem. Theory Comput.},
	author = {Kolodzeiski, Elena and Stein, Christopher J.},
	month = sep,
	year = {2023},
	pages = {6643--6655},
}

@article{pandharkar_localized_2021,
	title = {Localized {Active} {Space} {Pair}-{Density} {Functional} {Theory}},
	volume = {17},
	issn = {1549-9618},
	url = {https://doi.org/10.1021/acs.jctc.1c00067},
	doi = {10.1021/acs.jctc.1c00067},
	number = {5},
	urldate = {2026-07-22},
	journal = {J. Chem. Theory Comput.},
	publisher = {American Chemical Society},
	author = {Pandharkar, Riddhish and Hermes, Matthew R. and Cramer, Christopher J. and Truhlar, Donald G. and Gagliardi, Laura},
	month = may,
	year = {2021},
	pages = {2843--2851},
}

@article{hermes_variational_2020,
	title = {Variational {Localized} {Active} {Space} {Self}-{Consistent} {Field} {Method}},
	volume = {16},
	issn = {1549-9618},
	url = {https://doi.org/10.1021/acs.jctc.0c00222},
	doi = {10.1021/acs.jctc.0c00222},
	number = {8},
	urldate = {2026-07-22},
	journal = {J. Chem. Theory Comput.},
	publisher = {American Chemical Society},
	author = {Hermes, Matthew R. and Pandharkar, Riddhish and Gagliardi, Laura},
	month = aug,
	year = {2020},
	pages = {4923--4937},
}

@article{hermes_multiconfigurational_2019,
	title = {Multiconfigurational {Self}-{Consistent} {Field} {Theory} with {Density} {Matrix} {Embedding}: {The} {Localized} {Active} {Space} {Self}-{Consistent} {Field} {Method}},
	volume = {15},
	issn = {1549-9618},
	shorttitle = {Multiconfigurational {Self}-{Consistent} {Field} {Theory} with {Density} {Matrix} {Embedding}},
	url = {https://doi.org/10.1021/acs.jctc.8b01009},
	doi = {10.1021/acs.jctc.8b01009},
	number = {2},
	urldate = {2026-07-22},
	journal = {J. Chem. Theory Comput.},
	publisher = {American Chemical Society},
	author = {Hermes, Matthew R. and Gagliardi, Laura},
	month = feb,
	year = {2019},
	pages = {972--986},
}

@article{sand_multiconfiguration_2021,
	title = {A multiconfiguration pair-density functional theory-based approach to molecular junctions},
	volume = {155},
	issn = {0021-9606},
	url = {https://doi.org/10.1063/5.0063293},
	doi = {10.1063/5.0063293},
	number = {11},
	urldate = {2024-01-19},
	journal = {The Journal of Chemical Physics},
	author = {Sand, Andrew M. and Malme, Justin T. and Hoy, Erik P.},
	month = sep,
	year = {2021},
	pages = {114115},
}

@Misc{RUQT,
title = {Rowan University Quantum Transport, https://github.com/HoyLab-Rowan/RUQT},
}
%\begin{tocentry}
%\begin{center}
%\includegraphics[scale=1]{TOC_graph.eps}
%\end{center}
%\end{tocentry}

\end{document}